\documentclass[11pt, A4paper]{article}
\usepackage{caption}
\usepackage{subcaption}
\usepackage{amsmath}
\usepackage{verbatim}
\usepackage{setspace}
\usepackage{rotating}
\usepackage{morefloats}
\usepackage{hyperref}
\usepackage{siunitx}
\usepackage{pgf,tikz}
\usetikzlibrary{positioning}
\usepackage{graphicx, xcolor, color}
\usepackage{url}
\usepackage{amssymb}
\usepackage[sectionbib]{natbib}
\usepackage{sectsty}
\usepackage{multirow}
\usepackage{longtable}
\usepackage{rotating}
\usepackage[labelfont=bf,font=footnotesize]{caption}
\usepackage[left=2.56cm,right=2.56cm,top=2.56cm]{geometry}
\usepackage{mathrsfs}
\usepackage{dsfont}
\usepackage{enumerate,letltxmacro}
\usepackage[normalem]{ulem}
\LetLtxMacro\itemold\item

\usepackage[english]{babel}
\usepackage{blindtext}
\definecolor{MyGreen}{cmyk}{1.0,0.0,1.0,0.2}

\usepackage[font=singlespacing]{caption}

\begin{document}
\clearpage
\setcounter{page}{1}
\title{Alliances and Conflict, or Conflict and Alliances? Appraising the Causal Effect of Alliances on Conflict\footnote{The author is thankful to Jan Box-Steffensmeier, Bear Braumoeller, Skyler Cranmer, Weihua Li, and Giampiero Marra for their thoughtful comments.}}
\author{Benjamin W. Campbell\footnote{BWC: Doctoral Candidate in Political Science, The Ohio State University, e: \href{mailto:campbell.1721@osu.edu}{campbell.1721@osu.edu}}}
\date{\today}
\maketitle
\begin{center}Word Count:10,519 \\\end{center}

\begin{abstract} 
\noindent The deterrent effect of military alliances is well documented and widely accepted.  However, such work has typically assumed that alliances are exogenous.   This is problematic as alliances may simultaneously influence the probability of conflict and be influenced by the probability of conflict.  Failing to account for such endogeneity produces overly simplistic theories of alliance politics and barriers to identifying the causal effect of alliances on conflict.  In this manuscript, I propose a solution to this theoretical and empirical modeling challenge. Synthesizing theories of alliance formation and the alliance-conflict relationship, I innovate an endogenous theory of alliances and conflict.  I then test this theory using innovative generalized joint regression models that allow me to endogenize alliance formation on the causal path to conflict.  Once doing so, I ultimately find that alliances neither deter nor provoke aggression.  This has significant implications for our understanding of interstate conflict and alliance politics.  
\end{abstract}

\clearpage
\section{Introduction}
The study of military alliances, and in particular defensive military alliances, is central to the study of International Relations.  Indeed, the foundational works of IR examine the causes and consequences of formal and informal military alliances \citep{hans1948politics, waltz1979theory}.  Two central strands of this research assess the factors leading to the formation of military alliances and the effect of alliances on conflict.  The latter consists of a multiple threads considering the deterrent, provoking, empowering, or moderating effects of alliances on conflict.  This literaturie continues to garner attention, evidenced by a recent exchange over whether defensive alliance commitments truly deter aggression \citep{johnson2011defense, kenwick2015alliances, leeds2017theory, kenwick2017defense, morrow2017defensive}.  

While the literature on alliance formation has explicitly acknowledged the alliance-conflict relationship  \citep{morrow2000alliances}, the literature on the alliance-conflict relationship frequently neglects the literature on alliance formation \citep{tomz2015military, braumoeller2018flexible}.   Conventional wisdom argues that states, motivated by a common external security threat, aggregate their capabilities to ensure their security and survival.  If a state's security environment and expectation of conflict influences alliance formation, then treating an alliance as exogenous when assessing its deterrent or provocational effect makes little sense.  However, as a result of theoretical and empirical modeling decisions, scholars assume the expectation that states invoke an alliance does not influence their decision to form or maintain the alliance.  

This exogeneity assumption leads to misspecified theoretical and empirical models, producing inconsistent theoretical insights and barriers to causal identification.  
Failing to account for endogeneity \textit{will} produce incorrect inferences about the true effect of alliances \citep{pearl2009causality}.  
In other words, such problems have impacted prior findings on the alliance-conflict relationship, forcing us to question what we know about the deterrent or provoking effects of alliances \citep{braumoeller2018flexible}.

This manuscript grapples with foundational questions related to the literatures on alliance formation and the alliance-conflict relationship.  \textit{Do alliances deter aggression, even once endogenized? 
	Or does the expectation of aggression produce alliances?}  I argue that once synthesizing these literatures and producing an endogenous theory of alliances and conflict, our theoretical expectations about the effects of alliances change.  
Specifically, when linking the alliance formation and militarized dispute stages, it becomes apparent that scholars have omitted important aspects of threat, alliance reliability, and burden sharing.  
I expect a null effect of alliances on conflict once endogenizing alliances on the causal path to conflict and accounting for these important, unobserved confounders. 
These studies reveal that alliances do not appear to deter or provoke aggression, and instead, that conflict or the expectation of conflict may produce alliances.  
Significant improvements in model predictive performance confirm the robustness of these results.
My findings problematize the conventional wisdom on the deterrent effects of alliances, which has significant implications for how we consider deterrence, conflict, and alliance politics.

\section{Alliances and Conflict, or Conflict and Alliances?}

An extensive literature considers the alliance-conflict relationship.  While many have rigorously examined the effect of alliances on conflict, they have neglected its reverse, the effect of conflict on alliances.  
One cannot discount the role of external threats in motivating alliance formation, indicating a problematic omission in the alliance politics literature.

\subsection{Alliances and Conflict}
The theoretical and empirical literature on alliances and conflict is extensive and diverse.  There is both ongoing debate over whether alliances lead to or prevent conflict, and the mechanisms informing these relationships.  
Scholars have, at varying points, accepted alliances as both a cause of war and escalation \citep{scott1967functioning,  levy1981alliance} and as a cause of peace \citep{liska1962nations, singer1972capability}. 
Others have also highlighted the confusion regarding the typical direction of this relationship \citep{singer1972capability, levy1981alliance}.  
The inconsistency of these findings underscores the importance of an empirically rigorous, causal inference informed approach to studying the alliance-conflict relationship.    

\subsubsection{Alliances as Causes of Peace}
As mentioned, scholars continue to debate the direction of the alliance-conflict relationship \citep{leeds2003alliances, kenwick2015alliances, kenwick2017defense, leeds2017theory, morrow2017defensive}.  
The conventional wisdom, embodied by \citet{leeds2003alliances} and \citet{johnson2011defense}, argues that the presence of a relevant and explicit defensive alliance commitment deters aggression by a third party. 

A variety of mechanisms explain the potential pacifying effects of alliances.  
Those finding that alliances decrease conflict emphasize deterrence \citep{smith1995alliance, leeds2003alliances, johnson2011defense, leeds2017theory}, constraint \citep{snyder1997alliance, gelpi1999alliances, fang2014concede}, and information \citep{bearce2006alliances} as essential mechanisms. 
Deterrence has an intuitive logic -- ``If we assume that aggressors are more likely to initiate conflicts that they think they can win, and if we assume that usually aggressors are more optimistic about their ability to win a bilateral conflict than a multilateral conflict, it follows that potential aggressors should be more reluctant to challenge potential targets with allies committed to intervene on their behalf," \citep{leeds2003alliances}.  
Constraint holds that states may use alliances to constrain potentially bellicose allies.   
In these cases, allies may be able to credibly signal to the targeted state that if they refuse to concede to a demand (reasonable or not), the ally will not intervene on behalf of the targeted state \citep{fang2014concede}.  Finally, alliances may increase the information member-states have about others' military capabilities and reduce the information problems that lead to conflict between dyads near or at parity \citep{bearce2006alliances}.  

\subsubsection{Alliances as Causes of War}
An alternative perspective, while uncommon, holds that alliances can provoke and escalate conflict \citep{scott1967functioning,  levy1981alliance}.  In a well-known piece, \citet{levy1981alliance} argues that throughout much of history, war is followed by the formation of an alliance.  While he notes a variety of factors may undermine the generality of this conclusion, such as endogeneity, a pattern of behavior suggests a positive relationship between alliances and conflict.  
I will discuss these inferential challenges in greater detail later.

Three mechanisms help in understanding how alliances may promote or cause war.  Those finding that alliances induce conflict typically consider the role of the security dilemma \citep{snyder1984security}, entanglement \citep{beckley2015myth}, and moral hazard \citep{benson2013ally}.    
Within a rivalrous dyad, the formation of an alliance by one party may create a security dilemma, where in the increase in one state's security is perceived to decrease the security of another state, leading to arms races and crises \citep{snyder1984security}.    The second mechanism, entanglement, occurs when loyalty to an ally trumps self-interest and an ally that has no interest in intervening in their ally's dispute intervenes regardless.  Within this context, alliances have the opportunity to escalate isolated dyadic conflicts to general conflicts \citep{beckley2015myth}.  
Finally, alliances may create a moral hazard problem.
Revisionist states may become more bellicose and aggressive as they spread the costs of such behavior across their alliance partners (assuming support is unconditional) \citep{benson2013ally}.

\subsection{Conflict and Alliances}
Little work has attempted to disentangle the endogenous relationship between alliance formation and conflict initiation.  This seems problematic as scholars conventionally think alliances form in response to external threats \citep{hans1948politics, waltz1979theory, walt1985alliance, walt1990origins, morrow1991alliances, morrow2000alliances, mearsheimer2001tragedy, johnson2017external}, and patterns of conflict and threat may then go on to influence the interests of states and as such, their alliance decisions \citep{leeds2002alliance, gibler2006alliances, gibler2008costs,  braumoeller2013great}.  
To date, however, few have attempted to account for the possibility that alliances inform conflict and that conflict informs alliances.  
Treating this relationship as only unidirectional could create biased effect estimates and flawed inferences \citep{pearl2009causality, braumoeller2018flexible}.  To fill this gap, I consider this relationship as endogenous at the conventional dyadic levels.\footnote{A parallel analysis considers this relationship at the systemic level to provide greater understanding of this relationship.}

Since the founding of IR, scholars have invoked the balance of power to understand global politics -- that when threatened by the preponderant power of a potential rival, states aggregate their capabilities to increase their security \citep{hans1948politics, waltz1979theory, walt1990origins, mearsheimer2001tragedy}.  
Proponents of the CAM argue that when confronted by an external threat that undermines their security, states form alliances to institutionalize an aggregation of their resources \citep{morrow1991alliances, johnson2017external}. 

This proposition has empirical support.  States appear to form alliances as a response to their local security environment and in response to patterns of conflict that inform their interests  \citep{leeds2002alliance, gibler2006alliances, gibler2008costs, braumoeller2013great, johnson2017external}.  Indeed, many analyses have found the presence of a common enemy and the presence of a threatening adversary may drive a state to form alliances \citep{leeds2002alliance, gibler2006alliances, gibler2008costs, johnson2017external}.  One recent piece, which offers a refined theoretical and empirical approach to understanding the role of external threat in alliance formation, has found strong evidence that a threatened state will seek alliances \citep{johnson2017external}.  
In other words, the alliance formation literature indicates that the presence of a relevant alliance commitment is endogenous to the security environment of the member-states and their potential aggressors.  

Two problems when treating alliances commitments as an exogenous predictor of conflict.  First, treating alliances as exogenous to conflict, empirically or theoretically, runs contrary to the foundational theories of IR.  
This means that either the theoretical models used to argue that alliances influence conflict, or the empirical models used to assess alliances' effect on conflict are misspecified abstractions that do not conform to IR scholars' standard worldview \citep{walt1990origins, smith1995alliance}.  
Second, treating alliances and conflict as exogeneous when they are endogeneous will create biased estimates and flawed inferences \citep{smith1995alliance, pearl2009causality, braumoeller2018flexible}. 
There exist a variety of methodological opportunities to overcome this problem, including the design employed here: a dyadic approach using Generalized Joint Regression Models (GJRMs).  In the following section, I weave the strands of the alliance literature, providing a novel theoretical means to truly understand the endogeneity of alliances on the causal path to conflict.  

\section{Endogenizing the Alliance-Conflict Relationship}
Scholars widely accept that alliances reduce the risk of conflict.  
Unobservable features of states' security environments, however, drive both alliance formation and conflict.  
In other words, alliances, and the security environment prompting a state to form alliances, may be endogenous to conflict \citep{levy1981alliance, bearce2006alliances}.  Synthesizing theories of alliance formation and the alliance-conflict relationship offers an opportunity to truly understand the endogenous effect of alliances on conflict while clearly acknowledging the unobserved factors that confound this relationship.  

\begin{figure}
\centering
\scalebox{2}{
\begin{tikzpicture}
\node[text centered] (z) {$Z$};
\node[right = 1.5 of z, text centered] (x) {$X$};
\node[right=1.5 of x, text centered] (y) {$Y$};
\node[draw, rectangle, dashed, above = 1 of y, text centered] (u) {$U$};
 \node[draw, rectangle, above = 1 of x, text centered] (c) {$C$};
\draw[->, line width= 1] (z) --  (x);
\draw [->, line width= 1] (x) -- (y);
\draw[->, dashed, line width= 1] (u) --(x);
\draw[->, dashed, line width= 1] (u) -- (y);
\draw[->,line width= 1] (c) --(x);
\draw[->,line width= 1] (c) -- (y);
\end{tikzpicture}
}
\caption{\textbf{Directed Acyclic Graph (DAG) Illustrating Endogeneity.}  The causal effect of $X$ on $Y$ is unidentifiable as $U$ is some set of unobserved confounders producing endogeneity between $X$ and $Y$.}
\label{dag}
\end{figure}
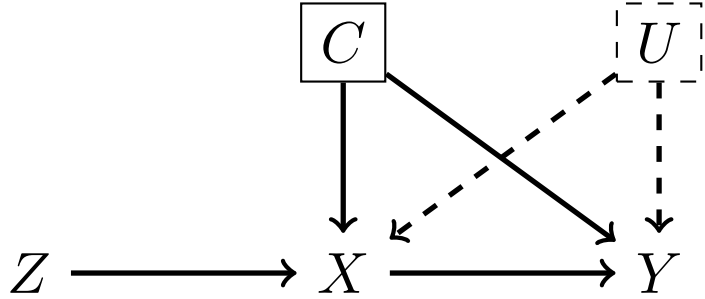

Consider the following, relatively simple Directed Acyclic Graph (DAG) \citep{pearl2009causality}.  
In Figure \ref{dag}, the causal effect of $X$ on $Y$, $X \rightarrow Y$ is the primary effect of interest.
$Z$ is some variable that fulfills the exclusion restriction, only influencing $Y$ through its effect on $X$.  Often times, a set of variables may influence the value of the treatment variable $X$ and the outcome $Y$.  These confounders are often both observed $C$, or unobserved $U$.  

Failing to account for $C$ and $U$ will lead to the estimation of a biased effect of $X \rightarrow Y$ as treatment assignment is non-random and there is some set of relevant omitted variables.  
Assuming no other confounders such as $U$ exist, conditioning upon $C$ and omitting $Z$ will correctly identify $X \rightarrow Y$.  
However, if present, some set of unobserved confounders $U$ will be correlated with $X$ and $Y$.  In this case, $U$ influences the values of $X$ making its values non-random while also influencing the values of $Y$.  
Failing to account for $U$ will produce endogeneity bias as an endogenous treatment $X$ is assigned non-randomly.  
In these cases, using $Z$ to model $X$ may allow for identification in certain models.\footnote{Note that endogeneity bias is theoretically and empirically distinct from sample selection bias, although the two are often conflated.  Endogeneity bias here means that there is some correlation between $X$ and other factors $U$ that influence the value of $X$, making its value non-random.  This becomes a problem when $U$ and $Y$ may also be correlated.  Sample selection bias, however, would refer to some process wherein the values of $X$ and $Y$ influence which observations are included in the sample.  This problem does not exist in this case as the sample reflects the census of all directed dyads during the time examined.}

?In the case presented here, the primary effect of interest is the causal effect of a targeted state's relevant defensive alliance commitments ($X$) on whether a target initiates a militarized dispute with the target ($Y$).  
A set of observable variables certainly influence alliance formation and conflict ($C$), such as the target and challenger's capabilities.  Assuming no other factors would influence this effect ($X \rightarrow Y$), simply conditioning upon $C$ would lead to correct inferences.  
However, through studying conflict we recognize that alliances ($X$) are not exogenous to conflict ($Y$), a variety of unobserved factors simultaneously influence the probability that states form and invoke an alliance through conflict ($U$). 
Scholars frequently acknowledge these factors, consider the simple case of alliance reliability, that when called upon to defend the targeted state, the ally might actually do so \citep{smith1995alliance, smith1996intervene, leeds2000reevaluating, leeds2003alliance}.  
Alliance reliability is simply one realization of the challenger and target's security environment and interests that lead us to conclude that alliances cannot be treated as exogenous to conflict.  
In addition, aspects of the threat posed by a potential aggressor may be unobserved confounders.  
As \citet{walt1985alliance, walt1990origins} notes, a variety of material and immaterial forces, often difficult to measure, constitute a state's threat perceptions.
The interest of a potential ally in sharing the burden of a potential alliance may also be an unobserved feature driving both alliance formation and conflict.  In the following section, I outline the empirical strategy used to identify the effect of alliances on conflict, including the operationalization and measurement of $Z$, $X$, $Y$, and $C$.  

The effect of alliances on conflict can be understood as a two-stage triadic process.  
In the first stage, $B$ surveys their security environment and interprets signals or indices made by $A$, making a decision to form an alliance with $C$ that would be relevant in a dispute with $A$, or not.  Once $B$ and $C$ form the alliance, or $B$ has abstained, the terminal stage begins as $A$ updates and reexamines their approach with $B$, either initiating a dispute with $B$ or being effectively deterred.  Linking these stages sheds light on the strategic decision-making of $A$, and how endogenzing alliance formation may influence our theoretical expectations about the effects of alliances on conflict.  

\subsection{First Stage: Alliance Formation}\label{1ststage}
In the first stage, $B$ responds to the threat of $A$ by ether forming an alliance with $C$ or abstaining.\footnote{This is, of course, a significant simplification of pre-conflict dynamics.  It is a pedagogically useful simplification, however.}  
The conventional model of alliances would predict that for state $B$ to form an alliance with $C$, the level of threat posted by $A$ much be sufficiently large such that the gain of alliance formation would outweigh its contracting and autonomy costs \citep{morrow1991alliances, morrow2000alliances}.  
The threat that $A$ poses to $B$ is a function of their aggregate power, proximity, offensive capability, and offensive intentions \citep{walt1985alliance, walt1990origins}.  To simplify this, threat can be constituted by a series of material and immaterial forces.  These material forces, such as aggregate power, proximity, and offensive capability are relatively easy to measure and account for in quantitative models of alliance formation.  Offensive intention, however, is immaterial and remains illusive.  

\citet{johnson2017external} notes that in measuring threat as the number of previous militarized disputes within a dyad over the last 5 or 10 years \citep{lai2000democracy, gibler2006alliances}, scholars neglect pre-dispute crisis bargaining dynamics.  Instead, he argues for measuring threat as the probability that a challenger $A$ would win a potential dispute, operationalizing threat as the challenger's capabilities and military expenditures relative to the target.  In other words, conventional measures of threat had tried to capture offensive intention while neglecting the other aspects of threat, such as $A$'s aggregate power, proximity to $B$, or offensive capability.  More recent measures, such as those employed in this manuscript and \citet{johnson2017external} have prioritized aggregate power and offensive capability.  These recent developments improve upon prior measurements as they more closely capture the aspect of threat they attempt to measure, even if the concept as a whole is left unmeasured.    

Unfortunately, measuring aggregate power, proximity, and offensive capability without capturing offensive intention reduces threat to teeth without intent to bite.  Alternatively, measuring offensive intention without capturing aggregate power, proximity, or offensive capability reduces threat to a toothless bite.  In other words, each of these components represents an essential component part of threat \citep{walt1985alliance, walt1990origins}.  While we have relatively strong measures of aggregate power, proximity, and offensive capability, we struggle in measuring offensive intention as it relies upon immaterial forces.  While we can attempt to capture intention using the number of prior disputes within a dyad over some arbitrary number of years, such measures will always be problematic as they rest upon immaterial, highly dynamic, and unmeasurable forces such as the $B$'s perceived resolve of $A$'s leaders.  

Beyond threat, a series of measurable and immeasurable factors may influence $B$'s decision to form an alliance.   The civil conflict history of $B$ may influence whether they seek alliance partners to help with domestic disputes \citep{schroeder1976alliances}.  This is dynamic is easily measurable.  Alternatively, the perception that $A$ will be effectively deterred by an alliance or if there are suitable and reliable allies for $B$ are not easily measured.  

\subsection{Second Stage: Conflict Initiation}\label{2ndstage}
In the second stage, $A$ updates, reexamining its relationship with $B$ given $B$'s alliance decisions and either initiates a dispute with $B$ or not.  The conventional model of deterrence holds that $A$ should be effectively deterred if the expected utility of winning is outweighed by the expected utility of losing \citep{hans1948politics, waltz1979theory, morrow1994alliances, smith1995alliance, smith1998extended}.  While this represents just one view on the causes for conflict, it is the view that has been dominant in the literature on the effect of alliances on conflict \citep{leeds2003alliances, johnson2011defense}.  

Similar to the first stage, many of the material and immaterial aspects of threat may influence the $A$'s expected utility from initiating a dispute with $B$.  As $A$'s aggregate power increases, the probability that they realize the gains of a dispute also increases as the resources they can devote to the war effort presumably scale \citep{mansfield1992concentration}.  The proximity of $A$ to $B$ influences the ability of $A$ to efficiently and effectively mobilize resources against $B$, influencing $A$'s expected utility through influencing the initial and prolonged costs of the dispute, and the probability that they win the dispute \citep{bueno1981war}.  The particular offensive capabilities of $A$ also influence their ability to wage war, and in particular, overcome $B$'s defenses \citep{jervis1976perception}.  Immaterial aspects of threat, namely offensive intention, influence the likelihood that $A$ attacks $B$ by defining the extractable gains, the selectorate's support for the dispute, or the resolve of $A$'s leaders \citep{bueno1981war}.  If $A$ has nothing to gain from a dispute, or no interest in initiating one, the costs of a dispute would outweigh the gains.  

Beyond threat, many of the same observable and unobservable factors that may inform $B$'s decision to form an alliance influence if $A$ decides to initiate a dispute with $B$.  For example, $A$ may be more (or less) likely to initiate a dispute against $B$ based upon foreign policy preference (dis)similarity, or if $A$ percieves $B$ to be domestically unstable \citep{schroeder1976alliances}.  The same unobserved features may also matter, if $B$ forms an alliance and $A$ lacks resolve, then $A$ may quickly deescalate the crisis.  Additionally, the potential deterrent benefits of an alliance partner $C$ may not be fully realized if $A$ perceives $C$ to be an unreliable ally.  

\subsection{Linking the Stages}
The alliance formation and conflict initiation stages are inexplicably linked, and treating them as distinct and alliance formation as exogenous produces overly simplistic theory and potential inferential errors.  They are linked by threat, and the broader security and crisis bargaining context.  Operationally, this is comprised by a series of unobserved confounders that nullify the effect of alliances on conflict.  The following discussion of this linkage provides a novel theoretical contribution by indicating how the endogenization of alliances matters, bringing to light new theoretical dynamics and empirical challenges.  In particular, I highlight the effects of threat, alliance reliability, and burden sharing as unobserved confounders that must be accounted for to ensure causal identification of the effect of alliances on conflict.  

\subsubsection{Threat and Offensive Intention as an Unobserved Confounder}
Consistent through the preceding discussion is the role of threat, broadly comprised by the material (aggregate power, proximity, offensive capabilities) and immaterial (offensive intention) forces that influence $B$'s perception of $A$, in alliance formation and conflict.  In Section \ref{1ststage}, threat was examined in its effects on alliance formation.  In order for $B$ to feel sufficiently threatened by $A$ such that seek out costly alliances, there must be the perception that $A$ has offensive intentions to use its aggregate power, the offensive capabilities to harm $B$, and that it is not constrained by distance \citep{walt1985alliance, walt1990origins}.  While measuring some aspects of this, such as aggregate power, proximity, and offensive capabilities is possible \citep{johnson2017external}, the measurement of offensive intention and what comprises it has proven illusive \citep{lai2000democracy, gibler2006alliances, johnson2017external}.  Threat, both in is material and immaterial constituting factors, confounds our ability to identify the causal effect of alliances and conflict, as indicated in Section \ref{2ndstage}.  While we can account for the material factors of threat as observed confounders, the immaterial factors that are potentially the most important are unobserved confounders, indicating the need for endogenizing alliances when modeling conflict.  

$A$'s offensive intention influences whether $B$ seeks an alliance with $C$, and whether $A$ has an interest in attacking $B$ in the first place.  As such, failing to account for this means perhaps the most foundational component of threat is left as an omitted confounding variable.  When $A$ has offensive intent, they may still be effectively deterred by an alliance between $B$ and $C$, and accounting for the other aspects of threat may be sufficient.  Alternatively, when $A$ lacks an offensive intent but other aspects of threat are present, it will appear as if an alliance between $B$ and $C$ will deter $A$ even if $A$ had no prior intention of engaging in a dispute.  As such, it seems reasonable that when omitting offensive intention, a negative relationship between alliances and conflict will be detected.  However, once including accounting for offensive intention, it seems apparent that a null expect may actually be expected. 

\subsubsection{Alliance Reliability as an Unobserved Confounder}
The perceived reliability of an ally may also influence the decision of an aggressor to initiate conflict \citep{smith1995alliance, smith1996intervene}. If $A$ perceives $C$ to be an unreliable ally, then $A$ is unlikely to be deterred and if they were intending to initiating a dispute they can do so safely.  This perception may simultaneously influence whether $B$ forms an alliance with $C$ and if $A$ subsequently initiates a dispute with $B$ and $C$ \citep{mattes2012reputation}.  By endogenizing alliances and incorporating this unobserved confounding, a null relationship between alliances and conflict may be more likely.  Should $C$ be perceived as a reliable ally, then it is possible that $A$ will still be deterred from engaging in a dispute with $B$.  However, if $A$ perceives $C$ to be an unreliable ally, then the presence of an alliance commitment may not effectively deter conflict and a null effect may be uncovered.

While attempts have been made to measure ally reliability, it has been done in general, context agnostic terms such as examining the number of times an ally $C$ has not upheld their commitment over some period of time \citep{miller2003hypotheses, mattes2012reputation}.  While reputation may be an useful and measurable heuristic, much of the literature has acknowledged that the reliability of an ally is contingent upon a series of context-specific and unmeasurable factors, such as their stake in the potential conflict or domestic constraints \citep{morrow1994alliances, morrow2000alliances, mattes2012reputation}.  Another approach has considered the microfoundations of alliance reliability -- finding that alliances may make publics more likely to support interventions \citep{tomz2015military}.  Regardless of measurement, an empirical and theoretical strategy that can account for an ally's reliability as an unobserved confounder may be fruitful as it provides a more rigorous opportunity assess the effect of alliance reliability on alliance formation and conflict.

\subsubsection{Burden Sharing as an Unobserved Confounder}
As previously mentioned, alliances are costly \citep{morrow1991alliances, morrow2000alliances}, and they typically thought to form only in cases where they are the most needed or thought to be most useful.  Conventionally, this typically implies that alliances are more likely form in cases where they would be most reliable and most likely to effectively deter \citep{smith1995alliance, smith1996intervene}.  However, it could also imply that alliances would form in cases where burden sharing is particularly essential \citep{olson1966economic, oneal1990theory, oneal1994theory}.    This view of alliances, often applied when understanding the North Atlantic Treaty Organization (NATO), essentially holds that states may seek allies to share the burden of defense when they have common enemies.  While those that benefit most may be expected to pay the greatest burden, there are many junior partners that are included that may not have the same stake in the alliances as many others.  

If the $C$ has ``skin in the game" and actively contributes the collective defense, $A$ may be more likely to be deterred. $A$ may have a particular view of $C$ as a burden sharing ally willing to stake resources in $B$'s defense.  In these cases, the formation of alliances would be non-random and correlated with the probability that they would effectively deter conflict.  However, in some larger multiparty alliances, members may join without the expectation of contributing much to the collective defense.  As such, we may detect a null relationship between alliances and conflict as $A$ percieves $C$ as less likely to stake resources in the defense of $B$.  

The previous cases illustrate how the formation of alliances and their subsequent challenging via militarized dispute are inextricably linked.  Modeling this linkage explicitly provides an opportunity to, for the first time, identify the causal effect of alliances on conflict.  In the following section I present a research design that provides an opportunity to rigorously examine this relationship.

\section{Empirical Strategy}\label{empiricalstrategy}

To test my endogenous theory of alliances and conflict, I primarily rely upon a dyadic study mirroring prior work.  To test this theory, I must endogenize the formation of alliances and the effect of the presence of a commitment on conflict.  To do so, I choose a two-equation model, and in particular, a Generalized Joint Regression Model (GJRM).  This model allows the analyst to account for endogeneity bias without particularly strict functional form assumptions, and as such, is the perfect model for my purpose \citep{braumoeller2018flexible}.  In this section, I discuss the empirical strategy used to model this endogeneity in dyadic perspective.

For this study the unit of analysis is a directed-dyad year, where a challenger $A$ is deciding to initiate a dispute against a target $B$ at year $t$.  The temporal domain is 1816-2000, and the data set initially used was taken from the replication archive for \citet{johnson2011defense}.  From this data, I added other necessary variables.    The first equation models whether state $B$ has a relative defensive alliance commitment in $t$ while the second equation models whether $A$ initiates a dispute with $B$ in $t$.  To model this process, I use bivariate recursive probit GJRM \citep{braumoeller2018flexible}.  This approach offers improvements over matching approaches which estimate unbiased treatment effects without making functional form assumptions, but assume fully specification of all relevant confounders, and over Heckman models which estimate unbiased treatment effects under non-random sample selection, but requires strict functional form assumptions.  

\subsection{Modeling Alliance Formation}

The outcome of interest in the first equation is whether $B$ has a defensive alliance commitment at time $t$ that could be invoked to their defense in a dispute with $A$ \citep{leeds2002alliance}. This variable, treated as an independent variable in \citet{johnson2011defense} and an outcome in \citet{johnson2017external}.  The first outcome in a two equation model, which later becomes the endogeneous predictor of the second outcome, requires at least one variable (preferably multiple, preferably strong) that fulfills the exclusion restriction \citep{sovey2011instrumental, braumoeller2018flexible}. For a variable to fulfill the exclusion restriction, a variable must only influence an outcome through its effect on an independent variable.  In the analyses presented here, I utilize two introduced by \citet{kimball2010political}.  The first measures the infant mortality rate of $B$ at $t$.  Infant mortality rate, used as a proxy for public goods provision, has been used to predict alliance formation with great success \citep{kimball2010political}.  The idea is fairly intuitive, as a state faces increased demand for public goods provision, they may attempt to ``outsource" their security provision through forming alliances so that they can shift spending on guns to spending on butter \citep{kimball2010political}.  Infant morality rate should likely only influence the probability that $A$ attacks $B$ through its effect on alliances, civil conflict, or $B$'s capabilities (the latter two may be conditioned out at the second stage).  To date, I find no published evidence of a direct relationship between a state's infant mortality rate and its propensity to be attacked.  The second variable used is the change in infant mortality rate from $t-1$ to $t$, which shows how shifts in the demand or provision of public goods may influence alliance formation \citep{kimball2010political}.  

To account for any other effect that infant mortality rates (and changes therein) may have on conflict initiation, target gross domestic product (GDP) per capita and civil war history are used.  First, $B$'s per capita GDP per year is included, with data from \citet{gleditsch2002expanded}.  This variable is included as infant mortality rates and its change, the instrumental variables used in my analyses, may influence GDP and then go on to influence conflict in the next equation by undermining economic productivity and development \citep{bremer1992dangerous}.  I also include a variable capturing the number of years in the last five years that $B$ has experienced civil conflict \citep{gleditsch2002armed}.  Infant mortality may produce civil conflict as rebellion may occur when domestic conditions are dire, which may also influence interstate conflict behavior \citep{elbadawi2002much, walter2004does}.  

I also utilize a set of control variables to model the CAM and other factors thought to influence alliance formation \citep{johnson2017external}.  First, to measure the strength of the threat $A$ poses to $B$, I include the challenger's ($A$) likelihood of winning the dispute.  This is measured as the challengers Composite Index of National Capabilities (CINC) score divided by the sum of CINC scores for both the challenger and target \citep{singer1972capability, johnson2017external}.  I also include the difference between $A$ and $B$'s CINC scores such that large positive values indicate that $A$ is much stronger than $B$ and large negative values indicate that $B$ is much stronger than $A$.  This captures the directionality in capability asymmetries.  These power asymmetries may not matter much, however, if the challenger does not have the resources necessary to engage in a dispute.  To account for this dynamic, I include the $A$'s CINC score.  Fourth, the threat $A$ poses to $B$ may also be a function of the proximity of the two countries \citet{walt1990origins, johnson2017external}.  This is measured as the natural log of minimum distance between $A$ and $B$ \citep{gleditsch2001measuring}.  Fifth, to measure the similarity in interests between $A$ and $B$ which may influence $j$'s threat perceptions \citep{kenwick2015alliances, leeds2017theory}, I include a measure foreign policy similarity using S-scores  \citep{signorino1999tau}.  Sixth, temporal dependencies within a rivalrous dyad may inform whether $B$ forms an alliance  \citep{kenwick2015alliances, leeds2017theory, johnson2017external}. As such, a nonparametric smoothed spline for peace years is added.  

A variety of monadic characteristics may also influence $B$'s threat perceptions and possible alliance opportunities.  To capture $B$'s general threat perception, I include the sum of CINC scores for all countries $B$ has had a dispute with in the 5 years prior to an observation.  This variation on a commonly used measure of threat aims to capture not just the number of rivals that a target has, but the strength of those targets.  The number of states contiguous to the target may also influence both their number of threats, or the number of potential allies.  As such, the number of directly contiguous states is calculated \citep{stinnett2002correlates}.  This variable is variant per target per year.  The political similarity of any two countries is also thought to shape their probability of alliance formation as states of heterophilous regime types may be less trusting of each other, or have incompatible interests \citep{gibler2006alliances}.  To create a monadic measure that varies by target year, I calculate the distance in a target's regime score from the global mean by year \citep{marshall2002polity}.  





\subsection{Modeling Conflict}
The variables included in the second model I primarily draw from the models included in the exchange between \citet{kenwick2015alliances, kenwick2017defense} and \citet{johnson2011defense, leeds2017theory}.  In fact, each control included in these pieces is represented here. These variables are relatively standard in most models of conflict.  For this equation, the outcome of interest is whether challenger $A$ initiates a dispute with target $B$ in year $t$, with data taken from \citet{maoz2005dyadic}.  In this section, I review the model of conflict and the variables included.  

Naturally, the predictor of conflict for this model, and naturally the essential variable of this manuscript, is whether the target $B$ has a relevant defensive commitment at time $t$.  As the outcome in the first model, this is the endogenous predictor of conflict and peace.  Taken from \citet{leeds2002alliance}, this variable measures if the target has a defensive alliance commitment in year $t$ that would be relevant in a dispute with challenger $A$.   

I also utilize a set of control variables that have been utilized in similar pieces \citep{johnson2011defense, kenwick2015alliances, kenwick2017defense, leeds2017theory}.  Most of these variables are also included in the prior model of alliance formation, and as such, I will only review them briefly.  Drawing from the literature on the democratic peace, the first control variable included accounts for the similarity in the challengers' and targets' regimes which may influence their probability of engaging in conflict \citep{maoz1993normative, russett1994grasping}.  In particular, a dichotomous variable for whether both states within a directed dyad are democratic is included \citep{marshall2002polity}. This indicator takes on a value of one when the lowest regime score within the dyad is greater than or equal to six.  The log of capital-to-capital distances is also used as disputes are more likely to emerge and less costly with geographically proximal states \citep{lemke2001relevance}.  Third, states with similar foreign policy portfolios may also be less likely to engage in disputes as they have similar interests and potentially more mediators \citep{signorino1999tau}.  

In addition, I include a series of variables that measure the capabilities of states within a directed dyad.    The fourth control included is $A$'s probability of winning a dispute with $B$ at time $t$. Assuming adequate information, a challenger should be unlikely to initiate a dispute if they are likely to lose \citep{slantchev2003principle}.   I also use another version of this variable which captures the difference in capabilities between a challenger and a target, which may be useful in capturing disparity in resource.  As an alternative to the often used indicator for whether the challenger is a great power, I use the continuous CINC score for the challenger which should capture their strength in absolute terms \citep{dafoe2013democratic}.  

The seventh variable included captures how dependent the challenger is on the target for their total trade volume.  Researchers have routinely shown that trade, both dependence and economic exchange, influences whether two states engage in a militarized dispute \citep{keshk2004trade}.  This variable is defined as the total volume of trade between challenger $A$ and target $B$, and the total trade volume of the challenger $A$ \citep{gleditsch2002expanded}.  

To help in fulfilling the exclusion restriction, I also include the target's GDP per capita and their number of years they've experienced civil war in the last five years in the conflict equation.  Each of these may influence conflict for the reasons discussed in the previous section.  Finally, as temporal dependencies influence the probability that $A$ and $B$ fight, I include a nonparametric smoothed spline for peace years \citep{metzger2018getting}.  

\section{Results}
I find support for my theory, once accounting for the endogeneity of alliances on the causal path to conflict, alliances do not appear to influence conflict.  Support is robust across model specifications.\footnote{Support is also robust to the level of analysis, as the systemic study presented in the SI Appendix uncovers support consistent with my expectation.}  This evidence is contrary to the expectation of much of the modern work on military alliances \citep{leeds2003alliances, johnson2011defense, leeds2017theory}, and indicates the importance of having empirical models that accurately reflect the substantive dynamics they represent.  

Based upon the previously articulated theory, at the dyadic level we would expect the presence of a relevant defensive alliance commitment would neither excite nor pacify a challenger.  Using the empirical strategy presented in Section \ref{empiricalstrategy}, I appraise this expectation.  Among the models that best fit the data out-of-sample, I uncover a null effect for the defensive commitment variable.  This indicates that once correctly accounting for the endogeneity of alliances on the causal path to conflict, challengers are neither deterred from nor induced to attack a given target.

I rely upon theoretical expectation and the data to recommend the copula function used to characterize the relationship between the residuals for the alliance formation and conflict equations.  An increase in the threat $A$ poses to $B$ should increase the likelihood that $B$ forms an alliance with $C$ and imply that $A$ is more likely to attack $B$.  A \ang{0} or \and{180} rotation of a copula function would describe this dynamic and the positive relationship and negative relationship between some unobserved confounder $U$ and the $X \rightarrow Y$ relationship.  Conventionally, it is thought that an increase in the perceived reliability of an ally $C$ should increase the probability that $B$ forms an alliance with $C$ and decrease the likelihood that $A$ attacks $B$ \citep{smith1996intervene}.  That said, reliable allies can be more costly than many states can afford which may indicate that a $B$ is less likely to form an alliance with $C$ if $C$ is perceived to be reliable \citep{mattes2012reputation}.  The dynamic described by \citet{smith1996intervene} would imply that a \ang{90} or \ang{270} copula rotation would best fit the data.  Alternatively, if \citet{mattes2012reputation} is correct, then perhaps a  \ang{0} or \and{180} would be preferable.  A more stark conclusion may be made when considering burden sharing, wherein $B$ would be more likely to ally with $C$ when $C$ is perceived to be more dedicated to a particular alliance.  When this dedication to burden-sharing is high, $A$ may be deterred as $C$ may be perceived as willing to stake significant resources in $B$'s defense.  This would imply that a \ang{90} or \ang{270} copula rotation would best fit the data.  

Unfortunately, we are unaware of which omitted variables exist and how they influence $X \rightarrow Y$.  As such, a rigorous data-driven approach may provide the best insights.  When allowing the data to recommend the copula function, I Iteratively over all available copula functions and examine out-of-sample model fit.  For each possible copula implemented in the \texttt{gjrm} package, which contains a sufficient collection of copula functions, I trained a model specified with the variables listed in Section \ref{empiricalstrategy} on 70\% of all dyads included in the data and computed Precision-Recall Area Under the Curve (PR-AUC) for the remaining 30\%.  I use the same test and train samples for each copula function.  The data are not sampled across multiple folds.  PR-AUC is a statistic that describes how well a given classifier handles recall, also known as the quotient of true positives to all positives (in this case, predicts militarized disputes that actually occur), and precision, which is the ratio of true positives and false negatives to all events (in this case, predicts militarized disputes and non-disputes.  By plotting these against one another and calculating the area under this curve, one can understand the trade-offs between precision and recall.  The copula found to maximize out-of-sample PR-AUC would assist in producing the best model fit while avoiding overfitting.  PR-AUC, in particular, is desirable when working with rare events data, such as militarized disputes \citep{cranmer2017can, campbell2018triangulating}.  This measure is also used to compare the improvements made in model fit when accounting for the endogeneity of military alliances.

\begin{figure}
    \centering
    \includegraphics[width=1\textwidth]{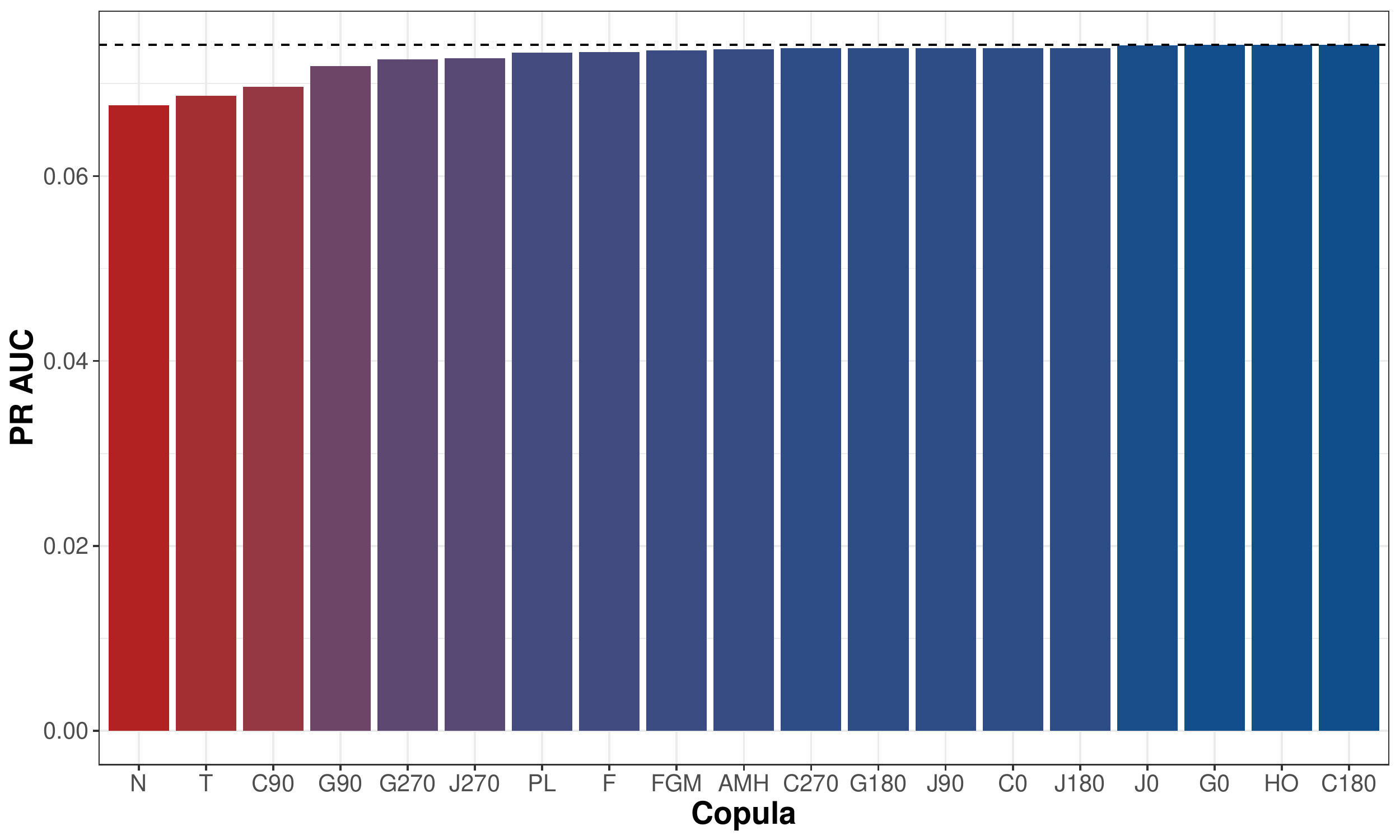}
    \caption{\textbf{Out-of-Sample Area Under the Precision-Recall Curve by Copula Function.}  The copulas producing the worst model fit are colored in red while the copulas producing the best model fit are colored in blue.  This is used to understand which GJRM models best fit predict conflict.  Dashed line refers to the maximum Precision-Recall area under the curve (PR AUC) achieved by the \ang{180} rotation of the Clayton copula function.  Naming of each function refers to the corresponding \texttt{GJRM} package labeling for each copula \citep{gjrmpack}.  }
    \label{copprauc}
\end{figure}

Figure \ref{copprauc} presents the result of this copula selection routine.  Some copulas, such as the \ang{180} rotation of the Clayton, Gumbel, or Hougaard copulas improve out-of-sample prediction relative to the Normal, Student-T, or \ang{90} rotation of the Clayton copulas.  The best fitting copula, the \ang{180} rotation of the Clayton distribution, represents a 9.6\% improvement in PR-AUC relative to the worst fitting copula, the Normal distribution.  While a nitty-gritty discussion of the relative merits of each copula are beyond the scope of this manuscript, some copulas better describe the relationship between the unobserved confounders and the alliance-conflict relationship and appear to better explain the relationship between the residuals for the alliance formation and conflict equations than others.  
As such, for the primary results presented in this section, the \ang{180} rotation of the Clayton distribution will be used as the copula.  This model would imply that for increases in any unobserved confounders, alliance formation and conflict should be more likely, capturing the expected influence of threat and alliance reliability \citep{mattes2012reputation}.  Fear not, the robustness of results to copula selection is later examined.

\begin{figure}
	\centering
	\includegraphics[width=1\textwidth]{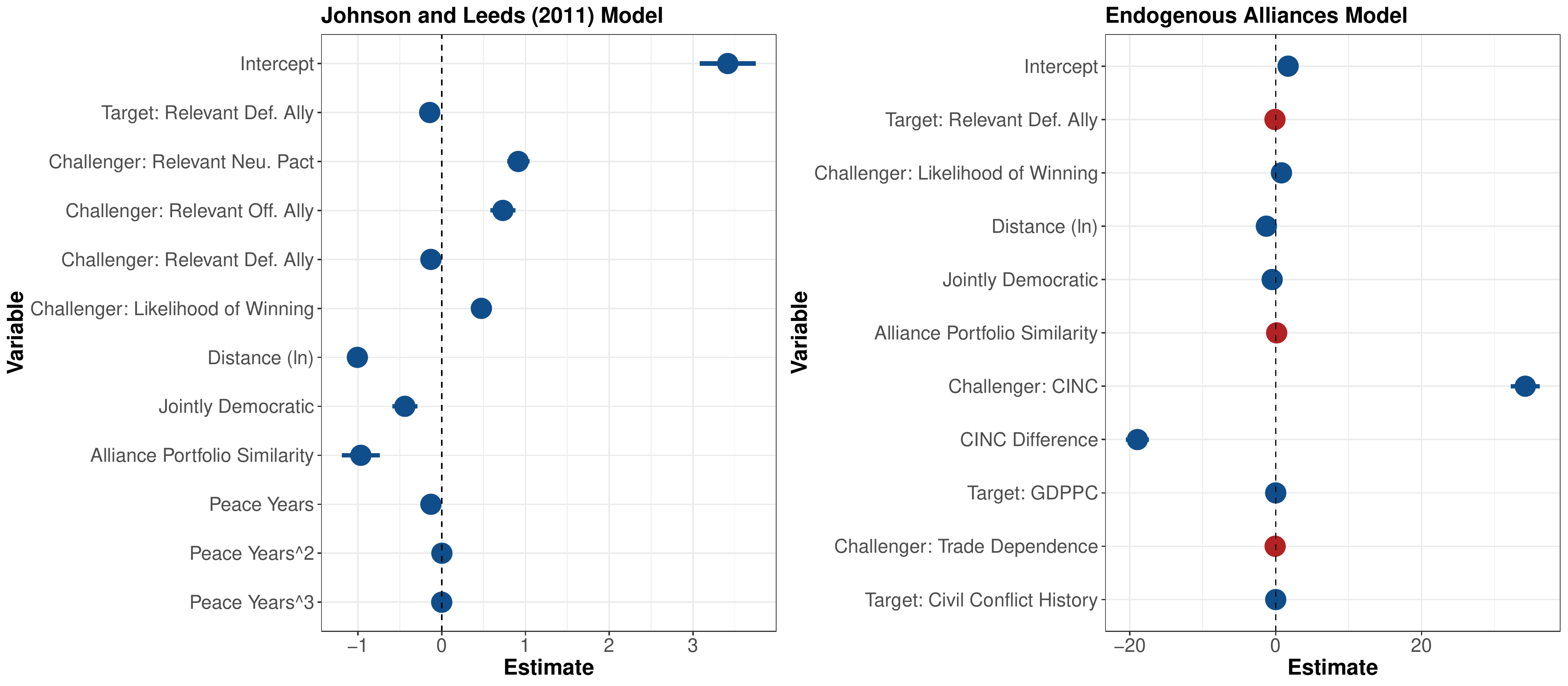}
	\caption{\textbf{Coefficient Plot for \citet{johnson2011defense} Model and Bivariate Model Conflict Equation.}  Bands refer to 95\% confidence intervals, while blue dots and bands refer to terms with confidence intervals excluding zero and red dots and bands refer to terms with confidence intervals including zero.  Endogenous alliances model includes a peace years spline.  This plot demonstrates that the baseline \citet{johnson2011defense} model uncovers the expected deterrent effect for defensive alliances, while the GJRM accounting for the endogeneity of the model does not uncover a deterrent effect.  The Average Treatment Effect (ATE) for the defensive alliance variable in the Endogenous Alliances Model is indistinguishable from zero at $-0.016 \,[-0.053, 0.016]$.}
	\label{coefplot}
\end{figure}

With a copula selected, a recursive bivariate probit Generalized Joint Regression Model is fit and interpreted.  I then compare these results to a logistic regression model of conflict replicated from \citet{johnson2011defense}.  The comparison to \citet{johnson2011defense} is fruitful as it is central to the study of alliances and conflict, spurring recent debate \citep{kenwick2015alliances, leeds2017theory, kenwick2017defense, morrow2017defensive}.  It also represents the current benchmark model of the alliance-conflict relationship.   Given the additional requirements of the bivariate model, I include a separate set of covariates.    Figure \ref{coefplot} presents the results from the primary bivariate model conflict equation and the logistic regression replication of \citet{johnson2011defense}.\footnote{Tests indicate that the two instrumental variables used, infant mortality and differenced infant mortality, are very strong instruments in predicting alliance formation.  $\chi^{2}$ analyses of the residual deviance and likelihood ratio tests for models with and without the instrument variables confirm that these are strong instruments at any conventional threshold.}   There represent significant differences in the inferences drawn from the model, particularly with the core variable of interest, whether the country targeted for a militarized dispute as a relevant defensive obligation that could be invoked in a dispute with the challenger.  In the results presented by \citet{johnson2011defense} and the mirroring replication here, a robust negative effect is detected for the relevant defensive obligation variable.  However, once endogenizing this variable effect is  indistinguishable from zero at any conventional $\alpha$ threshold.

I also calculate the average treatment (ATE) effect which is defined as the difference in outcomes under treatment and under control, in this case, what the effect of a target's relevant defensive commitment is on whether they are attacked by a challenger while accounting for all confounding variables.  For more details on this routine, the reader is directed to \citet{marra2011estimation}. Using 250 simulations from the posterior, I uncover an ATE of $-0.016$ with a 95\% credible interval straddling zero, $[-0.053,\,0.016]$.  This indicates that alliances do not have a statistically significant effect (at the $\alpha = 0.05$ level) on conflict independent of the circumstances that produce alliances and conflict.  Substantively, this means that once accounting for alliances as an endogenous variable, they only have a minuscule effect on conflict, on average only reducing the risk of attack by 1.6\%.  This effect is \textit{dramatically} smaller than the 20\% reduction in the predicted probability of being attacked for states having a relevant defensive obligation uncovered by \citet{johnson2011defense}.  While this, in and of itself, is not definitive evidence that alliances do not have an independent effect on militarized dispute onset, I will present additional evidence demonstrating such.  

The previous findings raise questions about which model best fits the data.  Demonstrating that endogenizing alliances leads to a null effect of alliances on conflict is meaningless if it does not lead to improvements in out-of-sample model fit.  Failing to see an improvement in model fit would likely indicate that the models have overfit the data and that the inferences drawn may be questionable.   As such, the predictions of previous models are assessed using PR Curves,  the previously described classification scheme best suited for militarized disputes data \citep{cranmer2017can}.  I begin with in-sample comparisons, moving to out-of-sample comparisons.  For the out-of-sample comparisons, the previously discussed models are trained on a random sample of the data consisting of 70\% of all observations.  The training model is then used to predict the remaining 30\% of observations. 

\begin{figure}
    \centering
    \includegraphics[width=1\textwidth]{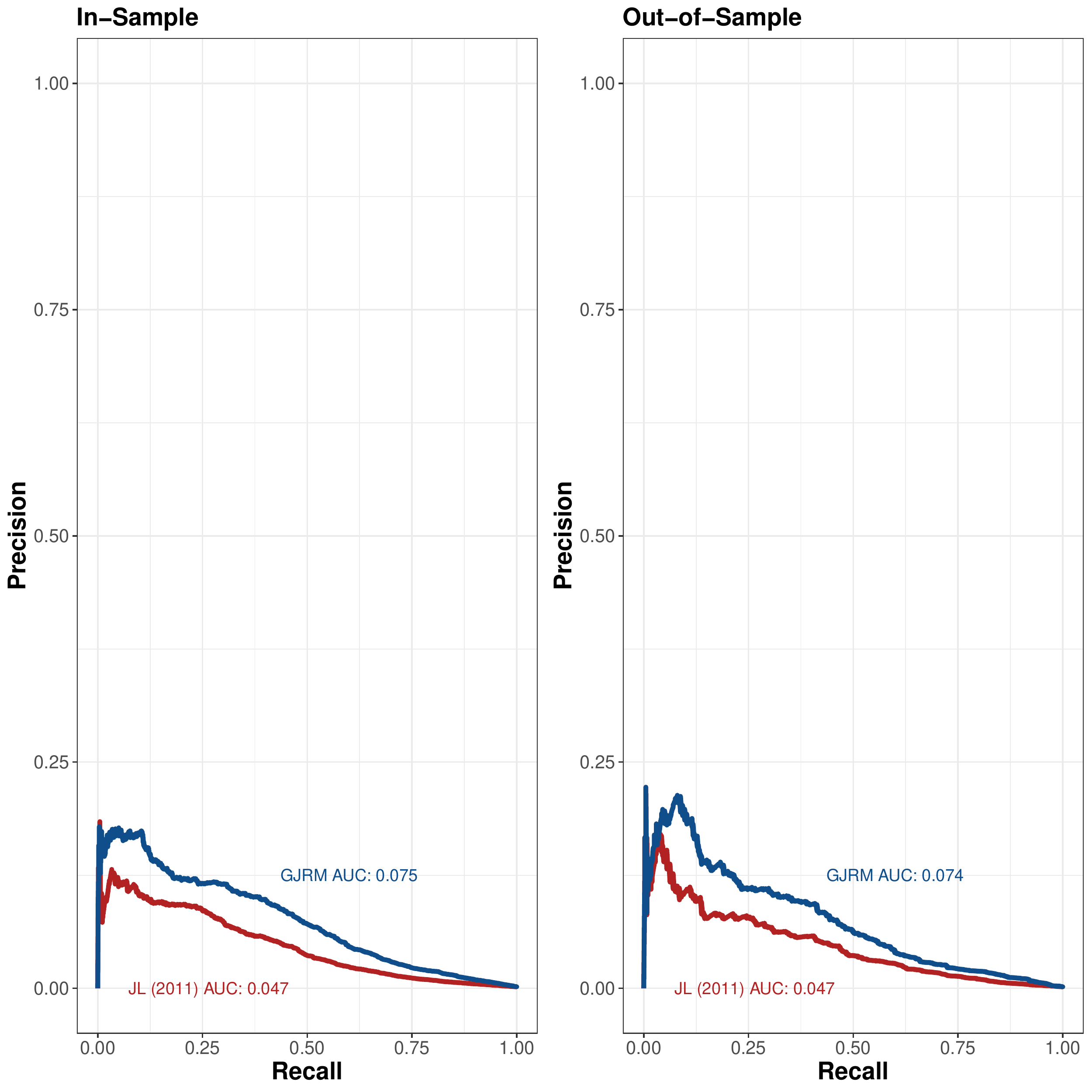}
    \caption{\textbf{Precision-Recall Curves for \citet{johnson2011defense} Model and Bivariate Model Conflict Equation.} Each plot is annotated with the curve's respective Precision-Recall area under the curve (PR AUC).  The GJRMs appear to present significant improvements in model fit.  The GJRM improves in-sample model fit by 59.6\% relative to the \citet{johnson2011defense} model. The GJRM improves out-of-sample model fit by 57.4\% relative to the \citet{johnson2011defense} model.  This indicates that by endogenizing alliances on the causal path to conflict, we can predict conflict more accurately by a large margin.  Both models do not do well in absolute terms as predicting conflict is relatively difficult.}
    \label{prcurves}
\end{figure}

With respect to both in-sample and out-of-sample prediction, the endogenous alliances model of conflict fairs better than the \citet{johnson2011defense} model that treats a target's relevant defensive alliance commitment as exogenous.  This is indicated by the PR Curves presented in Figure \ref{prcurves}.  The in-sample PR AUC for the endogenous alliances model is $0.075$, a 59.6\% improvement over the PR AUC for the exogenous alliances model.  When moving to out-of-sample prediction, PR AUC remains relatively constant, staying the same for the \citet{johnson2011defense} model while decreasing 1.4\% from $0.075$ to $0.074$ for the endogenous alliances model.  Regardless, however, the endogenous alliances model fits better out of sample than the baseline model.  

\begin{figure}
	\centering
	\includegraphics[width=1\textwidth]{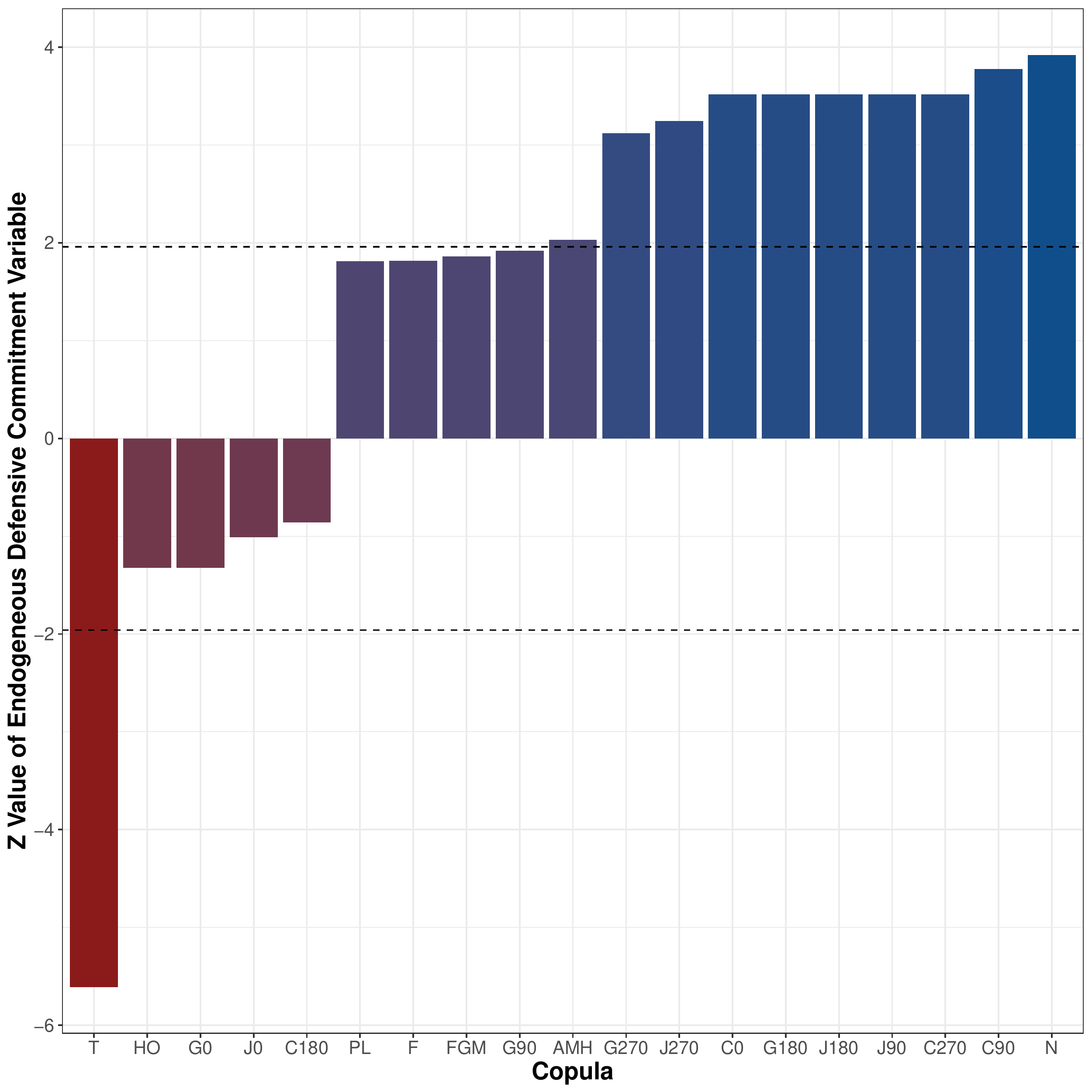}
	\caption{\textbf{Z Statistics for the Endogenous Relevant Defensive Alliance Commitment Variable in the Conflict Equation by Copula.} Bars are sorted and colored according to the value of the Z Statistic.  Dashed lines refer to z statistics that would be significant at the $\alpha = 0.05$ level.  Only one of 19 GJRM specifications fit yield the expected deterrent effect for alliances, while all others find a null or provocational effect.  This indicates that the null effect of defensive alliances on conflict is relatively robust.}
	\label{zstatbar}
\end{figure}

\begin{figure}
	\centering
	\includegraphics[width=1\textwidth]{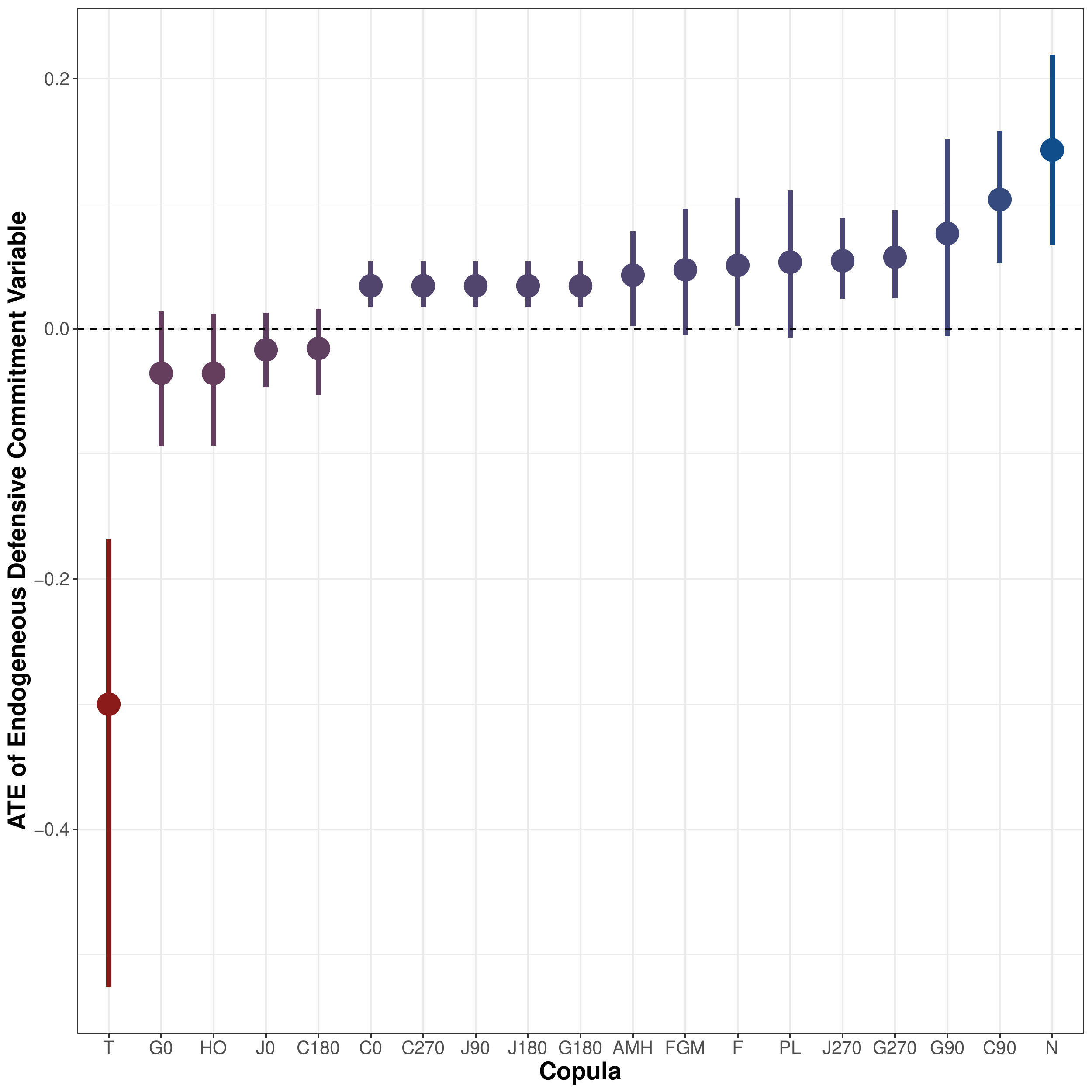}
	\caption{\textbf{Average Treatment Effects for the Endogenous Relevant Defensive Alliance Commitment Variable in the Conflict Equation by Copula.} Copulas are sorted and colored according to the value of the corresponding ATE.  ATEs reflect 250 iteration simulations from the posterior distribution.  An ATE that is not significant at the $\alpha = 0.05$ level will have a 95\% credible interval that contains zero, the dashed line.  Only one of 19 GJRM specifications fit yield the expected deterrent effect for alliances, while all others find a null or provocational effect.  This indicates that the null effect of defensive alliances on conflict is relatively robust.}
	\label{ateplots}
\end{figure}

To assess the relative robustness of the previously presented results, I conducted a sensitivity analysis that fit the previous endogenous alliances model on the full sample, iteratively changing the copula function used.  For each model, I examine the $z$-statistic for the effect of a target's relevant defensive alliance commitment on conflict.   The distribution of these $z$-statistics by copula, sorted by value, is presented in Figure \ref{zstatbar}.\footnote{Only the \ang{270} rotation of the Clayton produces a non-positive definite observed information matrix.   These positive definite matrices and relatively low largest absolute gradient values indicate that 18 of 19 models fit show evidence of convergence.}  Among the four copula found to best fit the data in Figure \ref{copprauc}, the \ang{180} rotation of the Clayton, Gumbel, Hougaard, and Joe, all would conclude a null effect for the endogenous alliance variable.  In fact, all but Student's-t, one of the worst fitting copulas, would lead the analyst to conclude that alliances either had a null effect or even a robust positive effect on conflict.  I uncover similar findings when examining the ATE for each copula, as indicated in Figure \ref{ateplots}.  These analysis indicated that once endogenizing alliances and accounting for all confounders, alliances do not deter conflict.

\section{Concluding Thoughts}
I have presented an endogenous theory of alliances and conflict which improves upon previous theories of the alliance-conflict relationship by linking the alliance formation and militarized dispute stages.  By synthesizing these literatures, I uncover unobserved confounders that make the formation of a relevant defensive alliance commitment endogenous to whether that commitment is invoked to defend a state in a militarized dispute.  In particular, the unobserved face of threat, offensive intention, the reliability of the ally, and the burden sharing dynamics of the alliance all influence alliance formation and the expectation that alliances will be invoked.  Previous theoretical models, while relatively parsimonious, miss an important part of the picture.  My approach makes significant improvements by treating alliances with care and in a way that does not contradict our theories of alliance formation.  
In the often utilized dyadic perspective, cutting-edge Generalized Joint Regression Models (GJRMs) uncover that once endogenizing alliances the presence of a relevant defensive alliance commitment neither deters nor provokes conflict.  

This research has three major implications.  First, and foremost, it represents a theoretical improvement over the standard practice of modeling the alliance-conflict relationship by being sensitive to the literature on alliance formation.  Given that the same unobserved factors that give rise to alliances also influence the presence of conflict, treating alliances as exogenous to conflict is problematic.  Second, this improvement is not just theoretical but it is also empirical as endogenizing conflict improves our ability to predict militarized disputes by 59.6\% in-sample and 57.4\% out-of-sample.  
Substantively, the difference is significant.  Previous models would expect that the presence of a relevant defensive alliance commitment would decrease the probability that a targeted state was attacked by 20\%.  When modeled correctly, I find that this deterrent effect drops from 20\% to a meager 1.6\%.  
Finally, this article may help to make greater sense for the inconsistent findings of the alliance-conflict literature and to shed light on recent debate \citep{johnson2011defense, kenwick2015alliances, leeds2017theory, kenwick2017defense, morrow2017defensive}.  Now that we have indicated, perhaps definitively, that alliances do not influence conflict, what are the moderating conditions that explain the inconsistent results for this relationship?  When do alliances deter aggression and when do alliances provoke aggression? Do the ally's intention or the target's lootability matter?  This finding must prompt researchers to better understand the nuanced relationship between alliance and conflict.   Each of these implications highlights the importance of being sensitive to how we treat alliances when modeling their effect on conflict.

\clearpage
\clearpage
\bibliographystyle{apsr}
\bibliography{AllianceEndogBib}

@article{campbell2018triangulating,
	Author = {Campbell, Benjamin and Cranmer, Skyler and Desmarais, Bruce},
	Journal = {arXiv preprint arXiv:1809.04141},
	Title = {Triangulating War: Network Structure and the Democratic Peace},
	Year = {2018}}

@article{tomz2015military,
	Author = {Tomz, Michael and Weeks, Jessica LP},
	Journal = {Unpublished manuscript, Stanford University},
	Title = {Military Alliances and Public Support for War},
	Year = {2015}}

@article{lemke2001relevance,
	Author = {Lemke, Douglas and Reed, William},
	Journal = {Journal of Conflict Resolution},
	Number = {1},
	Pages = {126--144},
	Publisher = {Sage Publications, Inc. 2455 Teller Road, Thousand Oaks, CA 91320},
	Title = {The relevance of politically relevant dyads},
	Volume = {45},
	Year = {2001}}

@article{oneal1994theory,
	Author = {Oneal, John R and Diehl, Paul F},
	Journal = {Political Research Quarterly},
	Number = {2},
	Pages = {373--396},
	Publisher = {Sage Publications Sage CA: Thousand Oaks, CA},
	Title = {The theory of collective action and NATO defense burdens: New empirical tests},
	Volume = {47},
	Year = {1994}}

@article{oneal1990theory,
	Author = {Oneal, John R},
	Journal = {International organization},
	Number = {3},
	Pages = {379--402},
	Publisher = {Cambridge University Press},
	Title = {The theory of collective action and burden sharing in NATO},
	Volume = {44},
	Year = {1990}}

@article{olson1966economic,
	Author = {Olson, Mancur and Zeckhauser, Richard},
	Journal = {The review of economics and statistics},
	Pages = {266--279},
	Publisher = {JSTOR},
	Title = {An economic theory of alliances},
	Year = {1966}}

@article{miller2003hypotheses,
	Author = {Miller, Gregory D},
	Journal = {Security studies},
	Number = {3},
	Pages = {40--78},
	Publisher = {Taylor \&amp; Francis},
	Title = {Hypotheses on reputation: alliance choices and the shadow of the past},
	Volume = {12},
	Year = {2003}}

@misc{bueno1981war,
	Author = {Bueno de Mesquita, Bruce},
	Publisher = {New Haven: Yale University Press},
	Title = {The war trap},
	Year = {1981}}

@article{mansfield1992concentration,
	Author = {Mansfield, Edward D},
	Journal = {Journal of Conflict Resolution},
	Number = {1},
	Pages = {3--24},
	Publisher = {Sage Publications 2455 Teller Road, Newbury Park, CA 91320},
	Title = {The concentration of capabilities and the onset of war},
	Volume = {36},
	Year = {1992}}

@article{walt1985alliance,
	Author = {Walt, Stephen M},
	Journal = {International security},
	Number = {4},
	Pages = {3--43},
	Publisher = {JSTOR},
	Title = {Alliance formation and the balance of world power},
	Volume = {9},
	Year = {1985}}

@article{lai2000democracy,
	Author = {Lai, Brian and Reiter, Dan},
	Journal = {Journal of Conflict Resolution},
	Number = {2},
	Pages = {203--227},
	Publisher = {Sage Publications, Inc. 2455 Teller Road, Thousand Oaks, CA 91320},
	Title = {Democracy, political similarity, and international alliances, 1816-1992},
	Volume = {44},
	Year = {2000}}

@article{mattes2012reputation,
	Author = {Mattes, Michaela},
	Journal = {International Organization},
	Number = {4},
	Pages = {679--707},
	Publisher = {Cambridge University Press},
	Title = {Reputation, symmetry, and alliance design},
	Volume = {66},
	Year = {2012}}

@article{smith1998extended,
	Author = {Smith, Alastair},
	Journal = {International Interactions},
	Number = {4},
	Pages = {315--343},
	Publisher = {Taylor \&amp; Francis},
	Title = {Extended deterrence and alliance formation},
	Volume = {24},
	Year = {1998}}

@article{morrow1994alliances,
	Author = {Morrow, James D},
	Journal = {Journal of Conflict Resolution},
	Number = {2},
	Pages = {270--297},
	Publisher = {Sage Periodicals Press 2455 Teller Road, Thousand Oaks, CA 91320},
	Title = {Alliances, credibility, and peacetime costs},
	Volume = {38},
	Year = {1994}}

@article{marra2011estimation,
	Author = {Marra, Giampiero and Radice, Rosalba},
	Journal = {Canadian Journal of Statistics},
	Number = {2},
	Pages = {259--279},
	Publisher = {Wiley Online Library},
	Title = {Estimation of a semiparametric recursive bivariate probit model in the presence of endogeneity},
	Volume = {39},
	Year = {2011}}

@manual{gjrmpack,
	Author = {Marra, Giampiero and Radice, Rosalba},
	Edition = {0.1-4},
	Month = {December},
	Organization = {Comprehensive R Archive Network},
	Title = {GJRM: Generalised Joint Regression Modelling},
	Year = {2017}}

@article{leeds2000reevaluating,
	Author = {Leeds, Brett Ashley and Long, Andrew G and Mitchell, Sara McLaughlin},
	Journal = {Journal of Conflict Resolution},
	Number = {5},
	Pages = {686--699},
	Publisher = {Sage Publications, Inc. 2455 Teller Road, Thousand Oaks, CA 91320},
	Title = {Reevaluating alliance reliability: Specific threats, specific promises},
	Volume = {44},
	Year = {2000}}

@article{leeds2003alliance,
	Author = {Leeds, Brett Ashley},
	Journal = {International Organization},
	Number = {4},
	Pages = {801--827},
	Publisher = {Cambridge University Press},
	Title = {Alliance reliability in times of war: Explaining state decisions to violate treaties},
	Volume = {57},
	Year = {2003}}

@article{smith1996intervene,
	Author = {Smith, Alastair},
	Journal = {Journal of Conflict Resolution},
	Number = {1},
	Pages = {16--40},
	Publisher = {Sage Periodicals Press 2455 Teller Road, Thousand Oaks, CA 91320},
	Title = {To intervene or not to intervene: A biased decision},
	Volume = {40},
	Year = {1996}}

@article{cranmer2017can,
	Author = {Cranmer, Skyler J and Desmarais, Bruce A},
	Journal = {Political Analysis},
	Number = {2},
	Pages = {145--166},
	Publisher = {Cambridge University Press},
	Title = {What Can We Learn from Predictive Modeling?},
	Volume = {25},
	Year = {2017}}

@unpublished{metzger2018getting,
	Author = {Metzger, Shawna K and Jones, Benjamin T},
	Title = {Getting Time Right: Using Cox Models and Probabilities to Interpret Binary Panel Data},
	Year = {2018}}

@article{keshk2004trade,
	Author = {Keshk, Omar MG and Pollins, Brian M and Reuveny, Rafael},
	Journal = {The Journal of Politics},
	Number = {4},
	Pages = {1155--1179},
	Publisher = {Cambridge University Press New York, USA},
	Title = {Trade still follows the flag: The primacy of politics in a simultaneous model of interdependence and armed conflict},
	Volume = {66},
	Year = {2004}}

@article{dafoe2013democratic,
	Author = {Dafoe, Allan and Oneal, John R and Russett, Bruce},
	Journal = {International Studies Quarterly},
	Number = {1},
	Pages = {201--214},
	Publisher = {Blackwell Publishing Ltd Oxford, UK},
	Title = {The democratic peace: Weighing the evidence and cautious inference},
	Volume = {57},
	Year = {2013}}

@book{russett1994grasping,
	Author = {Russett, Bruce},
	Publisher = {Princeton university press},
	Title = {Grasping the democratic peace: Principles for a post-Cold War world},
	Year = {1994}}

@article{maoz1993normative,
	Author = {Maoz, Zeev and Russett, Bruce},
	Journal = {American Political Science Review},
	Number = {3},
	Pages = {624--638},
	Publisher = {Cambridge University Press},
	Title = {Normative and structural causes of democratic peace, 1946--1986},
	Volume = {87},
	Year = {1993}}

@article{slantchev2003principle,
	Author = {Slantchev, Branislav L},
	Journal = {American Political Science Review},
	Number = {4},
	Pages = {621--632},
	Publisher = {Cambridge University Press},
	Title = {The principle of convergence in wartime negotiations},
	Volume = {97},
	Year = {2003}}

@misc{maoz2005dyadic,
	Author = {Maoz, Zeev},
	Title = {Dyadic MID dataset (version 2.0)},
	Year = {2005}}

@article{johnson2011defense,
	Author = {Johnson, Jesse C and Leeds, Brett Ashley},
	Journal = {Foreign Policy Analysis},
	Number = {1},
	Pages = {45--65},
	Publisher = {Oxford University Press},
	Title = {Defense pacts: A prescription for peace?},
	Volume = {7},
	Year = {2011}}

@article{elbadawi2002much,
	Author = {Elbadawi, Ibrahim and Sambanis, Nicholas},
	Journal = {Journal of conflict resolution},
	Number = {3},
	Pages = {307--334},
	Publisher = {Sage Publications Sage CA: Los Angeles, CA},
	Title = {How much war will we see? Explaining the prevalence of civil war},
	Volume = {46},
	Year = {2002}}

@article{walter2004does,
	Author = {Walter, Barbara F},
	Journal = {Journal of Peace Research},
	Number = {3},
	Pages = {371--388},
	Publisher = {Sage Publications Sage CA: Thousand Oaks, CA},
	Title = {Does conflict beget conflict? Explaining recurring civil war},
	Volume = {41},
	Year = {2004}}

@article{gleditsch2002armed,
	Author = {Gleditsch, Nils Petter and Wallensteen, Peter and Eriksson, Mikael and Sollenberg, Margareta and Strand, H{\aa}vard},
	Journal = {Journal of peace research},
	Number = {5},
	Pages = {615--637},
	Publisher = {Sage Publications London},
	Title = {Armed conflict 1946-2001: A new dataset},
	Volume = {39},
	Year = {2002}}

@article{bremer1992dangerous,
	Author = {Bremer, Stuart A},
	Journal = {Journal of Conflict Resolution},
	Number = {2},
	Pages = {309--341},
	Publisher = {Sage Publications 2455 Teller Road, Newbury Park, CA 91320},
	Title = {Dangerous dyads: Conditions affecting the likelihood of interstate war, 1816-1965},
	Volume = {36},
	Year = {1992}}

@article{gleditsch2002expanded,
	Author = {Gleditsch, Kristian Skrede},
	Journal = {Journal of Conflict Resolution},
	Number = {5},
	Pages = {712--724},
	Publisher = {Sage Publications Thousand Oaks},
	Title = {Expanded trade and GDP data},
	Volume = {46},
	Year = {2002}}

@article{marshall2002polity,
	Author = {Marshall, Monty G and Jaggers, Keith and Gurr, Ted Robert},
	Journal = {College Park: University of Maryland},
	Title = {Polity IV project: Dataset users' manual},
	Year = {2002}}

@article{stinnett2002correlates,
	Author = {Stinnett, Douglas M and Tir, Jaroslav and Diehl, Paul F and Schafer, Philip and Gochman, Charles},
	Journal = {Conflict Management and Peace Science},
	Number = {2},
	Pages = {59--67},
	Publisher = {Sage Publications Sage CA: Thousand Oaks, CA},
	Title = {The correlates of war (cow) project direct contiguity data, version 3.0},
	Volume = {19},
	Year = {2002}}

@article{signorino1999tau,
	Author = {Signorino, Curtis S and Ritter, Jeffrey M},
	Journal = {International Studies Quarterly},
	Number = {1},
	Pages = {115--144},
	Publisher = {Wiley Online Library},
	Title = {Tau-b or not tau-b: measuring the similarity of foreign policy positions},
	Volume = {43},
	Year = {1999}}

@article{gleditsch2001measuring,
	Author = {Gleditsch, Kristian S and Ward, Michael D},
	Journal = {Journal of Peace Research},
	Number = {6},
	Pages = {739--758},
	Publisher = {Sage Publications/PRIO 6 Bonhill Street, London EC2A 4PU, UK.},
	Title = {Measuring space: A minimum-distance database and applications to international studies},
	Volume = {38},
	Year = {2001}}

@article{kimball2010political,
	Author = {Kimball, Anessa L},
	Journal = {Journal of Peace Research},
	Number = {4},
	Pages = {407--419},
	Publisher = {Sage Publications Sage UK: London, England},
	Title = {Political survival, policy distribution, and alliance formation},
	Volume = {47},
	Year = {2010}}

@article{sovey2011instrumental,
	Author = {Sovey, Allison J and Green, Donald P},
	Journal = {American Journal of Political Science},
	Number = {1},
	Pages = {188--200},
	Publisher = {Wiley Online Library},
	Title = {Instrumental variables estimation in political science: A readers' guide},
	Volume = {55},
	Year = {2011}}

@article{braumoeller2018flexible,
	Author = {Braumoeller, Bear F and Marra, Giampiero and Radice, Rosalba and Bradshaw, Aisha E},
	Journal = {Political Analysis},
	Number = {1},
	Pages = {54--71},
	Publisher = {Cambridge University Press},
	Title = {Flexible causal inference for political science},
	Volume = {26},
	Year = {2018}}

@article{morrow2000alliances,
	Author = {Morrow, James D},
	Journal = {Annual Review of Political Science},
	Number = {1},
	Pages = {63--83},
	Publisher = {Annual Reviews 4139 El Camino Way, PO Box 10139, Palo Alto, CA 94303-0139, USA},
	Title = {Alliances: Why write them down?},
	Volume = {3},
	Year = {2000}}

@book{pearl2009causality,
	Author = {Pearl, Judea},
	Publisher = {Cambridge university press},
	Title = {Causality},
	Year = {2009}}

@article{gibler2008costs,
	Author = {Gibler, Douglas M},
	Journal = {Journal of Conflict Resolution},
	Number = {3},
	Pages = {426--454},
	Publisher = {Sage Publications Sage CA: Los Angeles, CA},
	Title = {The costs of reneging: Reputation and alliance formation},
	Volume = {52},
	Year = {2008}}

@article{gibler2006alliances,
	Author = {Gibler, Douglas M and Wolford, Scott},
	Journal = {Journal of Conflict Resolution},
	Number = {1},
	Pages = {129--153},
	Publisher = {Sage Publications Sage CA: Thousand Oaks, CA},
	Title = {Alliances, then democracy: An examination of the relationship between regime type and alliance formation},
	Volume = {50},
	Year = {2006}}

@article{johnson2017external,
	Author = {Johnson, Jesse C},
	Journal = {International Studies Quarterly},
	Pages = {sqw054},
	Publisher = {Oxford University Press},
	Title = {External Threat and Alliance Formation},
	Year = {2017}}

@article{morrow2017defensive,
	Author = {Morrow, James D},
	Journal = {The Journal of Politics},
	Number = {1},
	Pages = {341--345},
	Publisher = {University of Chicago Press Chicago, IL},
	Title = {When Do Defensive Alliances Provoke Rather than Deter?},
	Volume = {79},
	Year = {2017}}

@article{kenwick2017defense,
	Author = {Kenwick, Michael R and Vasquez, John A},
	Journal = {The Journal of Politics},
	Number = {1},
	Pages = {329--334},
	Publisher = {University of Chicago Press Chicago, IL},
	Title = {Defense Pacts and Deterrence: Caveat Emptor},
	Volume = {79},
	Year = {2017}}

@book{jervis1976perception,
	Author = {Jervis, Robert},
	Publisher = {Princeton University Press},
	Title = {Perception and misperception in international politics},
	Year = {1976}}

@article{morrow1991alliances,
	Author = {Morrow, James D},
	Journal = {American Journal of Political Science},
	Pages = {904--933},
	Publisher = {JSTOR},
	Title = {Alliances and asymmetry: An alternative to the capability aggregation model of alliances},
	Year = {1991}}

@article{schroeder1976alliances,
	Author = {Schroeder, PW},
	Journal = {Knorr (ed.),(Lawrence, Kan., 1976)},
	Pages = {227--56},
	Title = {Alliances, 1815-1945: Weapons of Power and Tools of Management', Historical Dimensions of National Security Problems, K},
	Year = {1976}}

@book{mearsheimer2001tragedy,
	Author = {Mearsheimer, John J},
	Publisher = {WW Norton \&amp; Company},
	Title = {The tragedy of great power politics},
	Year = {2001}}

@article{hans1948politics,
	Author = {Morgenthau, Hans},
	Journal = {The Struggle for Power and Peace. Third Edition, Translated by M. Starkova, New York, USA},
	Title = {Politics Among Nations},
	Year = {1948}}

@article{leeds2002alliance,
	Author = {Leeds, Brett and Ritter, Jeffrey and Mitchell, Sara and Long, Andrew},
	Journal = {International Interactions},
	Number = {3},
	Pages = {237--260},
	Publisher = {Taylor \&amp; Francis},
	Title = {Alliance treaty obligations and provisions, 1815-1944},
	Volume = {28},
	Year = {2002}}

@book{walt1990origins,
	Author = {Walt, Stephen M},
	Publisher = {Cornell University Press},
	Title = {The origins of alliance},
	Year = {1990}}

@book{braumoeller2013great,
	Author = {Braumoeller, Bear F},
	Publisher = {Cambridge University Press},
	Title = {The great powers and the international system: systemic theory in empirical perspective},
	Year = {2013}}

@book{waltz1979theory,
	Author = {Waltz, Kenneth},
	Publisher = {Reading, Mass.: Addison-Webley},
	Title = {Theory of international relations},
	Year = {1979}}

@article{benson2013ally,
	Author = {Benson, Brett V and Bentley, Patrick R and Ray, James Lee},
	Journal = {Journal of Peace Research},
	Number = {1},
	Pages = {47--58},
	Publisher = {Sage Publications Sage UK: London, England},
	Title = {Ally provocateur: Why allies do not always behave},
	Volume = {50},
	Year = {2013}}

@book{snyder1997alliance,
	Author = {Snyder, Glenn H},
	Publisher = {Cornell University Press},
	Title = {Alliance politics},
	Year = {1997}}

@article{gelpi1999alliances,
	Author = {Gelpi, Christopher},
	Journal = {Imperfect unions: Security institutions over time and space},
	Pages = {107--39},
	Publisher = {Oxford University Press Oxford},
	Title = {Alliances as instruments of intra-allied control},
	Year = {1999}}

@article{smith1995alliance,
	Author = {Smith, Alastair},
	Journal = {International Studies Quarterly},
	Number = {4},
	Pages = {405--425},
	Publisher = {The Oxford University Press},
	Title = {Alliance formation and war},
	Volume = {39},
	Year = {1995}}

@article{snyder1984security,
	Author = {Snyder, Glenn H},
	Journal = {World politics},
	Number = {04},
	Pages = {461--495},
	Publisher = {Cambridge Univ Press},
	Title = {The security dilemma in alliance politics},
	Volume = {36},
	Year = {1984}}

@article{leeds2017theory,
	Author = {Leeds, Brett Ashley and Johnson, Jesse C},
	Journal = {The Journal of Politics},
	Number = {1},
	Pages = {000--000},
	Publisher = {University of Chicago Press Chicago, IL},
	Title = {Theory, Data, and Deterrence: A Response to Kenwick, Vasquez, and Powers},
	Volume = {79},
	Year = {2017}}

@article{kenwick2015alliances,
	Author = {Kenwick, Michael R and Vasquez, John A and Powers, Matthew A},
	Journal = {The Journal of Politics},
	Number = {4},
	Pages = {943--954},
	Publisher = {University of Chicago Press Chicago, IL},
	Title = {Do Alliances Really Deter?},
	Volume = {77},
	Year = {2015}}

@article{leeds2003alliances,
	Author = {Leeds, Brett Ashley},
	Journal = {American Journal of Political Science},
	Number = {3},
	Pages = {427--439},
	Publisher = {Wiley Online Library},
	Title = {Do alliances deter aggression? The influence of military alliances on the initiation of militarized interstate disputes},
	Volume = {47},
	Year = {2003}}

@article{fang2014concede,
	Author = {Fang, Songying and Johnson, Jesse C and Leeds, Brett Ashley},
	Journal = {International Organization},
	Number = {04},
	Pages = {775--809},
	Publisher = {Cambridge Univ Press},
	Title = {To concede or to resist? The restraining effect of military alliances},
	Volume = {68},
	Year = {2014}}

@article{beckley2015myth,
	Author = {Beckley, Michael},
	Journal = {International Security},
	Number = {4},
	Pages = {7--48},
	Publisher = {MIT Press},
	Title = {The myth of entangling alliances: Reassessing the security risks of US defense pacts},
	Volume = {39},
	Year = {2015}}

@article{bearce2006alliances,
	Author = {Bearce, David H and Flanagan, Kristen M and Floros, Katharine M},
	Journal = {International Organization},
	Number = {03},
	Pages = {595--625},
	Publisher = {Cambridge Univ Press},
	Title = {Alliances, internal information, and military conflict among member-states},
	Volume = {60},
	Year = {2006}}

@book{liska1962nations,
	Author = {Liska, George},
	Publisher = {Johns Hopkins Press},
	Title = {Nations in alliance: The limits of interdependence},
	Year = {1962}}

@article{singer1972capability,
	Author = {Singer, J David and Bremer, Stuart and Stuckey, John},
	Journal = {Peace, war, and numbers},
	Pages = {48},
	Title = {Capability distribution, uncertainty, and major power war, 1820-1965},
	Volume = {19},
	Year = {1972}}

@book{scott1967functioning,
	Author = {Scott, Andrew MacKay},
	Publisher = {New York: Macmillan},
	Title = {The functioning of the international political system},
	Year = {1967}}

@article{levy1981alliance,
	Author = {Levy, Jack S},
	Journal = {Journal of Conflict Resolution},
	Pages = {581--613},
	Publisher = {JSTOR},
	Title = {Alliance formation and war behavior: An analysis of the great powers, 1495-1975},
	Year = {1981}}

\end{document}